\documentclass[twocolumn,aps,showpacs,preprintnumbers,amsmath,amssymb, superscriptaddress]{revtex4-1}
\pdfoutput=1

\usepackage{bm}
\usepackage{graphicx}
\usepackage{color,xspace}

\renewcommand{\vec}{\mathbf}

\begin{document}

\title{Inelastic electron tunneling spectroscopy of nanoporous gold films}

\author{H. W. Liu}\email{hliu@iphy.ac.cn}
\affiliation{Institute of Physics, Chinese Academy of Sciences, Beijing 100190, China}
\author{R. Nishitani}
\affiliation{Department of Basic Sciences, Kyushu Institute of Technology, Fukuoka 804-8550, Japan}
\author{T. Fujita}
\affiliation{WPI-Advanced Institute for Materials Research, Tohoku University, Sendai 980-8577, Japan}
\affiliation{PRESTO, JST, 4-1-8 Honcho Kawaguchi, Saitama 332-0012, Japan}
\author{W. Li}
\affiliation{Department of Physics, Tsinghua University, Beijing 100084, China}
\author{L. Zhang}
\affiliation{WPI-Advanced Institute for Materials Research, Tohoku University, Sendai 980-8577, Japan}
\author{X. Y. Lang}
\affiliation{WPI-Advanced Institute for Materials Research, Tohoku University, Sendai 980-8577, Japan}
\author{P. Richard}
\affiliation{Institute of Physics, Chinese Academy of Sciences, Beijing 100190, China}
\author{K. S. Nakayama}
\affiliation{WPI-Advanced Institute for Materials Research, Tohoku University, Sendai 980-8577, Japan}
\author{X. Chen}
\affiliation{Department of Physics, Tsinghua University, Beijing 100084, China}
\author{M. W. Chen}\email{mwchen@wpi-aimr.tohoku.ac.jp}
\affiliation{WPI-Advanced Institute for Materials Research, Tohoku University, Sendai 980-8577, Japan}
\author{Q. K. Xue}
\affiliation{Department of Physics, Tsinghua University, Beijing 100084, China}
\affiliation{WPI-Advanced Institute for Materials Research, Tohoku University, Sendai 980-8577, Japan}

\date{\today}

\begin{abstract}
We investigated the localized electronic properties of nanoporous gold films by using an ultra-high vacuum scanning tunneling microscope at low temperature (4.2 K). Second derivative scanning tunneling spectroscopy shows the plasmon peaks of the nanoporous gold films, which are excited by inelastic tunneling electrons. We propose that the nanorod model is appropriate for nanoporous gold studies at the nanometer-scale. These results are supported by a 3D electron tomography analysis and theoretical calculations of nanoporous gold with ellipsoid shape.
\end{abstract}



\maketitle

\section{Introduction}
At the core of plasmonics, a surface plasmon is a coherent electron oscillation that is excited by an electromagnetic field and propagates along a metal-dielectric interface, which is now widely used in molecular biology, environmental monitoring, and energy harvesting and conversion \cite{Polman_Science322}. With a three-dimensional (3D) continuous network structure of interconnecting gold ligaments and nanopore channels, recently-developed dealloyed nanoporous gold (NPG) films are very active plasmonic materials \cite{F_Yu_AnalChem78, Y_Ding_MRS34}. In particular, a smaller pore size leads to a larger enhancement in surface enhanced Raman scattering (SERS) \cite{LH_Qian_APL90}. Such films have been proved to be outstanding free-standing SERS substrates able to achieve single molecule detection \cite{HW_Liu_SciRep1}. Yet, although easy and controllable synthesis allows to tune the plasmonics and SERS properties of NPG films, there is currently no complete theoretical understanding of these properties due to the complex morphology of the films \cite{XY_Lang_JPCC113}. A dealloyed nanoporous metallic film is characterized by various kinds of nanostructures, including nanopores and nanoligaments, where the plasmons are confined. Strong electromagnetic fields appear due to the large curvatures of these nanofeatures and to the short distance between neighbor ligaments \cite{XY_Lang_JPCC113}. Practical knowledge of inelastic electron tunneling spectroscopy (IETS)  tells us that the optical constant of metals \cite{Berndt_PRL27}, as well as the size, shape, assembly of nanostructures \cite{TZ_Han_JJAP48} and scanning tunneling microscopy (STM) tip-sample cavity \cite{HW_Liu_PRB79} can affect the plasmonic behavior of gold films. One may ask: are the local electronic properties of nanofeatures in a nanoporous film more similar to those of nanoclusters or nanorods? This understanding is necessary when optimizing the performances of NPG films and designing NPG-based nano-optics devices. Unfortunately, literature lacks of quantitative approach to the local electro-optical properties of such nanoporous material. 

IETS is an appropriate experimental technique for investigation of low-lying excitations in a wide variety of experimental systems in the few eV's energy range \cite{Adkins_JPC_Sol18}. It uses quantum-mechanical tunneling of electrons through thin insulating barriers in solid-state structures. A plasmon occurs in a bulk metal at a frequency $\omega_p$ given by $\omega_p^2 = ne^2/\varepsilon_0\varepsilon_rm_e$, where $n$ is the electron density. Electron energy-loss experiments show that plasmons are easily excited by the passage of energetic electrons. Characteristic values of $\omega_p$ for bulk metals fall in the $5\sim10$ eV range, outside the range of IETS. However, by using a degenerate semiconductor, $n$ can be reduced enough to shift the energy of the plasma oscillations below 1 eV. In addition, surface plasmons occur near the insulator-metal interface (\emph{i.e.} at the sample surface), with electric fields decaying exponentially into both metal and insulator with frequencies often smaller than $\omega_p$. In metals, such excitation of surface plasmons can generally be inferred only from subsequent decay to visible radiation \cite{Mills1982}.

IETS measurements at the normal boiling point of $^4$He (4.2 K) can give an energy resolution as good as 2 meV. To benefit of its high sensitivity, in this work we explore the local electronic properties of NPG films by using IETS, and show that at the nanometer-scale the nanorod features of the NPG films rather than nanoclusters play a dominant role in determining the plasmonics properties of these materials.

\section{Experiment}

NPG films were synthesized by chemically etching Ag$_{65}$Au$_{35}$ (atomic ratio) leaves with a size of $\sim 20$ mm $\times$ 20 mm $\times$ 100 nm in a 70\% (mass ratio) HNO$_3$ solution at room temperature. Nanopore sizes are controlled by the corrosion time \cite{LH_Qian_APL90,XY_Lang_JPCC113}. The intermediate porous structure was quenched by distilled water and residual acid within nanopore channels was removed by water rinsing. The nanopore sizes, defined by the equivalent diameters of nanopore channels or gold ligaments \cite{Fujita_APL92}, are determined using a rotational fast Fourier transform method \cite{Fujita_JJAP47}. The films are physically supported by highly ordered pyrolytic graphite substrates, which do not have any plasmon peak in the IETS measurement range. 

The local structure of nanoporous gold was characterized by scanning electron microscopy (SEM) (JEOL, JIB-4600F) and spherical-aberration-corrected transmission electron microscopy (STEM)(JEOL, JEM-2100F). The topography of the NPG films was measured using a JEOL STM at ambient condition. As-prepared NPG films were transferred into a low-temperature ultra-high vacuum scanning tunneling microscope (LT UHV-STM, UNISOKU) for IETS measurements [see Fig. \ref{Fig1_V_topography}(a)]. Each IETS spectrum (proportional to the second derivative of the I-V characteristic) was measured at liquid helium temperature (4.2 K) with the standard lock-in method \cite{YS_Fu_PRL103}. The modulation voltage of the excitation signal for the measurements was 7 mV. The I-V curves were measured below 3.0 eV for tip stability.

\section{Results and discussion}

In 1976, J. Lambe and S. L. McCarthy discovered light emission from inelastic electron tunneling when a current was drawn through a metal-oxide-metal tunneling junction \cite{Lambe_PRL923}. This technique yields a broad-band light source with a high frequency linear cutoff that depends only upon the applied voltage, and a peak in $d^2I/dV^2$ of a Ag junction was observed for bias coinciding with the plasmonic frequency $\nu_{0}$ (\emph{i. e.} when $|eV| = h\nu_{0}$). The effect can be interpreted in terms of inelastic tunneling excitation of optically coupled surface plasmon modes. Later, light emission from condensed Ag films as intense as $10^{-3}$ photon per injected electron has been observed in a STM \cite{Coombs_JMicros325, Gimzewski_EuroPhysLett435}, in perfect agreement with a theoretical work by Persson and Baratoff \cite{Persson_PRL3224} who also found that hot-electron injection yields a much smaller estimate, of the order of $\sim 10^{-6}$. In other words, if the energy of the tunneling electrons is larger than the energy of the dipole active surface plasmon mode, most of the emitted light results from inelastic tunneling, in agreement with the STM studies.

\begin{figure}[!t]
\begin{center}
\includegraphics[width=3.4in]{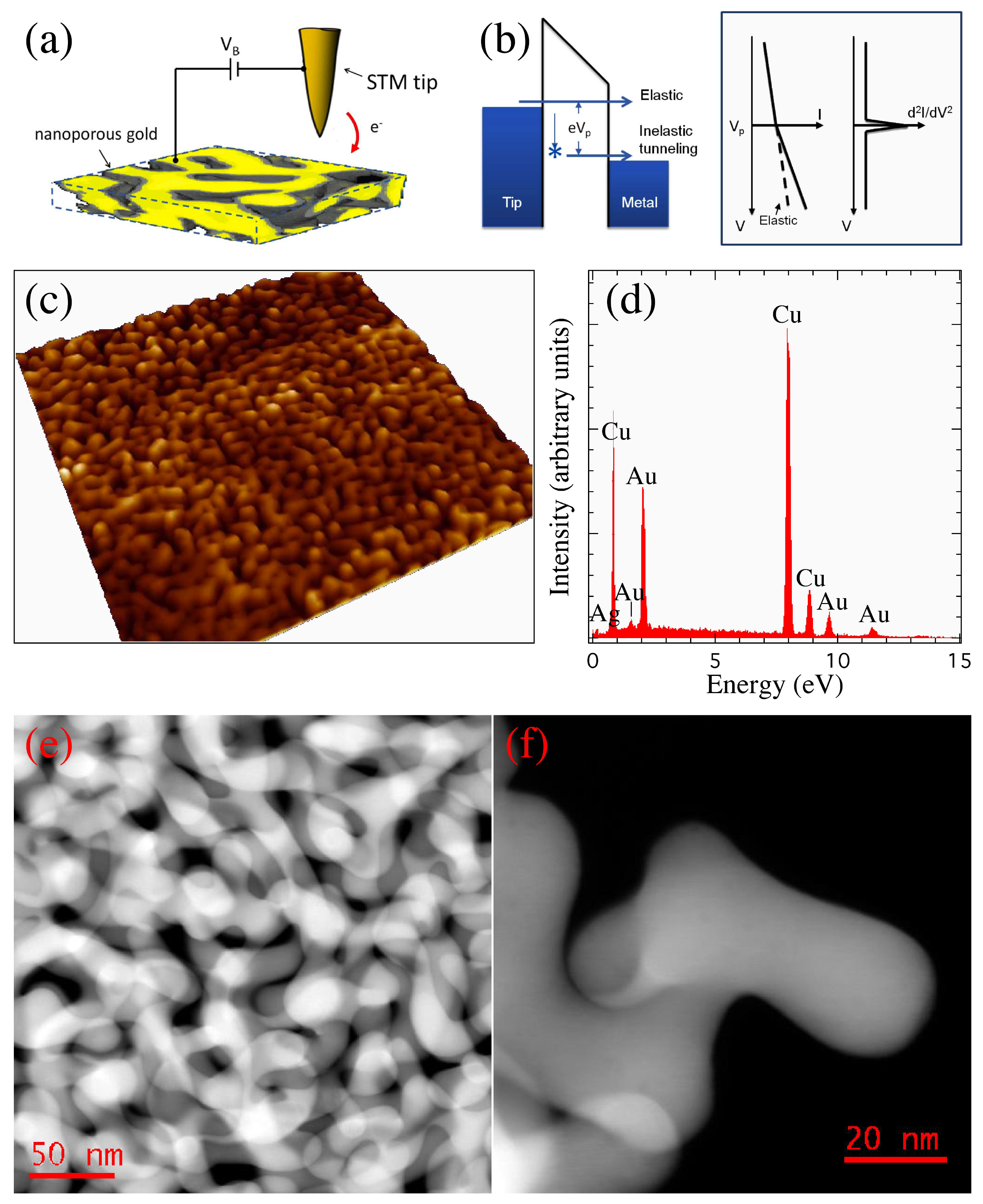}
\end{center}
\caption{\label{Fig1_V_topography}(Color online). (a) Schematic representation of inelastic tunneling microscopy from a NPG film in a STM. (b) Illustration of inelastic tunneling spectroscopy. (c) A typical STM topographic image of a NPG film. The scan size is $2\times 2$ $\mu$m$^2$. $V_B = -1.0$ V , $I_t = 1.0$ nA. (d) EDX spectrum of a dealloyed NPG film. STEM images of non-annealed NPG film containing $\sim 5$ at. \% residual Ag in (e) and detailed image in (f).}
\end{figure}

As illustrated in Fig. \ref{Fig1_V_topography}(b), the inelastic interactions of electrons tunneling from a STM tip to the sample with a characteristic energy loss mode $\omega = eV_p$ open up an additional tunneling channel on top of the elastic tunneling current background when the bias voltage reaches $V_p$, which leads to an increase in the I-V curve slope. The second derivative of the I-V curve thus exhibits a peak corresponding to the slope change in the I-V curve and the area below the peak is directly related to the strength of the interaction \cite{Adkins_JPC_Sol18}. The nature of the plasmons is determined by the boundary conditions of potentials of the STM tip and the metal surface \cite{Umeno_PRB54}, and the spectra are thus closely related to the geometric structure of the STM tip and the metal being sampled. 

A typical STM image in Fig. \ref{Fig1_V_topography}(c) illustrates gold nanoaggregates on the surface. The peak-peak distance for 20 nm pore size NPG films is about 70 nm. The conductivity and flatness of the NPG films indicate that they are suitable candidates for STM studies. The concentration of residual silver in the dealloyed NPG films detected by SEM energy dispersive x-ray (EDX) analysis is smaller than 3 at.\% [see Fig. \ref{Fig1_V_topography}(d)]. The Cu features in the EDX analysis are attributed to the copper stage of the sample. STEM results show that the local atomic structure is homogeneous \cite{Fujita_NatMater775} and there is no observable Ag-rich region on the surface of non-annealed samples when the residual Ag concentration is smaller than 5\% [see Fig. \ref{Fig1_V_topography}(e) and \ref{Fig1_V_topography}(f)], and the residual Ag mainly distributes in the inner ligaments instead of on the NPG surface because of the lower electrochemical stability of Ag on the ligament surface \cite{Zhang_JPCC19583}. Theoretically, we considered the effect of Ag mixing on the optical properties (plasmon) by calculating the complex dielectric constant using the effective medium model of Maxwell-Garnet \cite{Niklasson_JAP1984}. The result shows that the complex dielectric constant is almost unchanged by small mixing of Ag (3\%).Therefore, the presence of Ag is ignored in our IETS measurements and analysis.

Jacobsen \emph{et al.} \cite{Jacobsen_ASS252} reported that there is a significant variation of the tunneling behavior in spatially-resolved spectroscopy from a nanoporous film of TiO$_{2}$. The authors attributed this energy shift to the quality difference in the aggregates interconnections in the nanoporous network. It is worth noticing that unlike in ordinary films, the visible features in the STM images, isolated or interconnected at the surface, are in fact connected to the deeper network. Therefore, they cannot be directly used to establish a correlation between the plasmonic resonances and the local geometric structures. To overcome this problem, our investigation includes a large amount of IETS spectra and conclusions are based on statistics. 

\begin{figure}[!t]
\begin{center}
\includegraphics[width=3.4in]{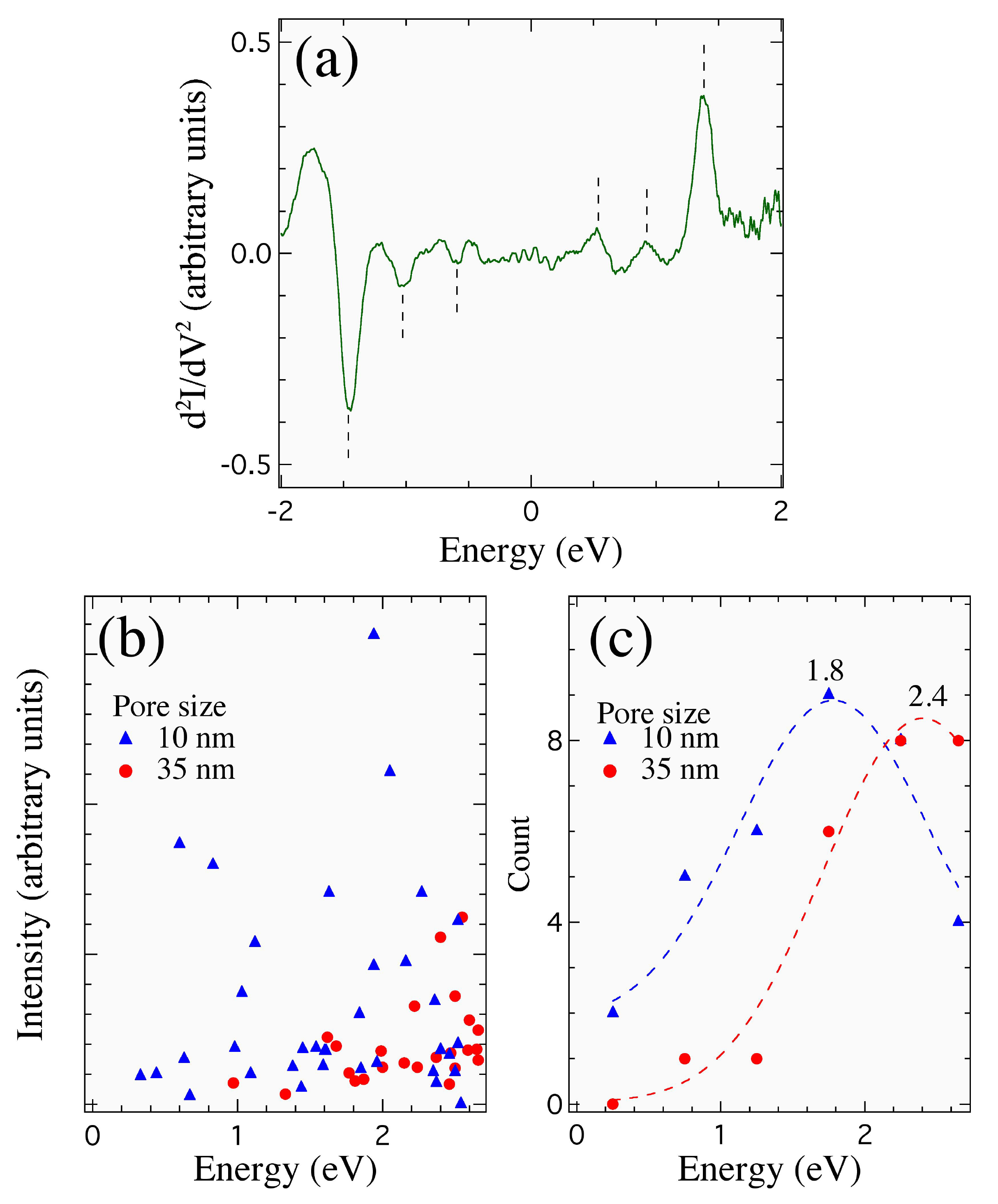}
\end{center}
\caption{\label{Fig2_V_IETS}(Color online). (a) Typical $d^2I/dV^2$ curve from a NPG film. Dashed lines are guides to the eye. (b) Energy and intensity distributions of the plasmon peaks in the $d^2I/dV^2$ curves for NPG films of 10 nm (blue) and 35 nm (red) pores. (c) Gaussian distributions (dashed lines) of the plasmon peaks in (c).}
\end{figure}

Here one should note that from our experimental data some NPG ligaments behaves like a bulk. Indeed, for some measured ligaments there is no peak in the second derivative curves within our measured range. This property may be caused by the complex connecting network structure. A typical second derivative spectrum of NPG film measured by STM for a ligament presenting features, and spanning a large energy range (from -2.0 V to 2.0 V) in both bias polarities, is shown in Fig. \ref{Fig2_V_IETS}(a). The main features in the $d^2I/dV^2$ curve are symmetrically located in energy with respect to the zero-bias position. Due to the symmetry they can be assigned to inelastic plasmon contributions.

For each sample, we took 30 IETS spectra from various ligaments. The two films show obvious differences. For the film with 10 nm diameter pore size, more abundant and sharper peaks frequently appear in the low energy range, down to 0.5 eV. In contrast, NPG films with larger ligaments (35 nm) are less actively excited in this range, as shown in Fig. \ref{Fig2_V_IETS}(b). Accordingly, the gaussian fits of the number of excitations per energy range peak around 1.8 eV and 2.4 eV for 10 nm and 35 nm pore sizes, respectively [see Fig. \ref{Fig2_V_IETS}(c)]. Considering that most excitations in the IETS spectra locate in the $0.5\sim$ 2.5 eV range, it is reasonable to assign their origin to elementary excitations of local plasmons at the surface of the NPG films \cite{Adkins_JPC_Sol18}. The variation in the peak position is mainly caused by the average sizes of the gold ligaments.

It is instructive at this stage to compare our results with previous data of plasmonic excitations reported on simpler gold-based nanostructures. Nilius \emph{et al.} \cite{Nilius_PRB65} studied plasmon excitations in single gold clusters using IETS. For the tip-sample coupling, the energy shift of the emission line can be assigned to a relative cluster-size effect on the tip-induced plasmons. In contrast to Mie-like plasmons at 2.3 eV, plasmon peak energies approximately decrease proportionally with the cluster size and can be fit by $\hbar\omega(d)=2.3\textrm{ eV}-0.02\textrm{ eV/nm}\times d(\textrm{nm})$. Instead of nanoclusters, Mohamed \emph{et al.} \cite{Mohamed_ChemPhysLett317} studied the lightning of 20 nm diameter gold nanorods with length ranging from 20 to 100 nm and found that the peak maxima are continuously red-shifted with increasing the aspect ratio of the nanorods. The optical absorption properties of Au nanoparticles in the visible range are very much dictated by the effect of the boundary conditions of the coherent electron oscillations and also due to the $d\rightarrow sp$ interband electronic transitions. S\"{o}nnichsen \emph{et al.} \cite{Sonnichsen_PRL88} studied single-particle scattering spectra from both nanospheres and nanorods. The spectra show the particle-plasmon resonances of the spheres at higher energy (2.19 eV) than the long-axis mode of the rods (1.82 eV).

\begin{figure}[!t]
\begin{center}
\includegraphics[width=3.4in]{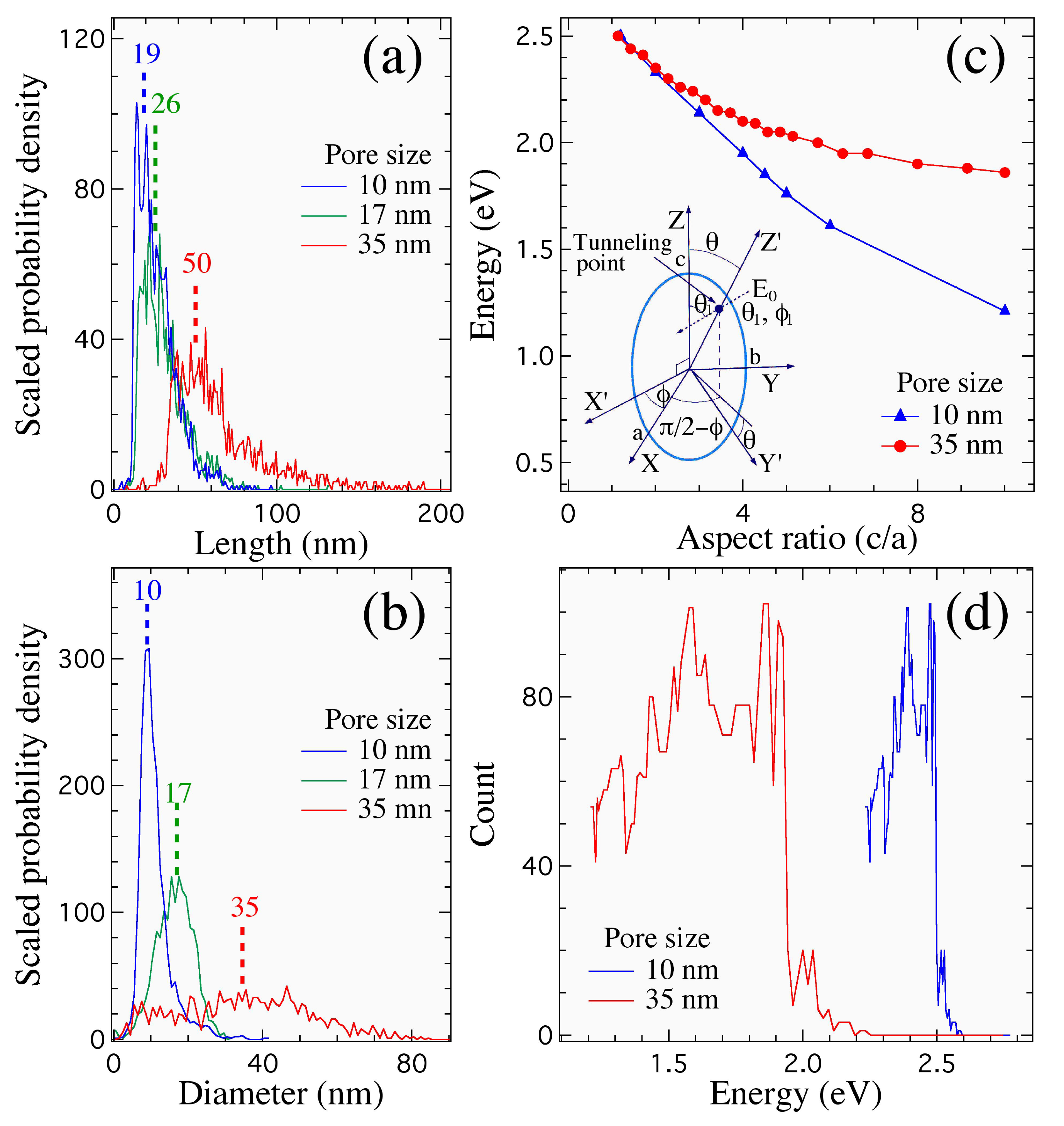}
\end{center}
\caption{\label{Fig3_V_TEM}(Color online). (a) Distribution of scaled probability density of inter-junction path length of gold ligaments, and (b) distribution of scaled probability density of the diameters of gold ligaments for the films with 10, 17, and 35 nm pore sizes. (c) (Inset: Schematic drawing of structural characteristics of NPG by an ellipsoid.) Calculated results of aspect ratio ($c/a$) dependence of plasmon energy for different samples with average particle sizes of 10 nm and 35 nm. (d) Conversion of abscissa axis from length in (a) to energy.}
\end{figure}

Our IETS results suggest that the plasmonic properties of NPG are more consistent with the nanorod approximation than the nanocluster model. To confirm this assumption, we characterized  the morphology of NPG films with different pore sizes using 3D electron tomography. To extract the length and diameter distributions of the nanorods, we translated the nanostructure into a skeletal network using a 3D thinning algorithm by which the volume of voxel data are digitally reduced from the surface, pixel by pixel, until skeletonized pixels remain at the center of the nanopore channels or gold ligaments \cite{Fujita_APL92}. Fig. \ref{Fig3_V_TEM}(a) displays the distributions of the path lengths of the gold ligaments in films corresponding to 5 min, 15 min, and 6 hours dealloying. The distributions are asymmetric and exhibit a broad maximum at $\sim19$, 26, and 50 nm, respectively. Fig. \ref{Fig3_V_TEM}(b) illustrates the diameter distributions of the gold ligaments, which show gaussian profiles peaked at 10, 17 and 35 nm, for 5 min, 15 min, and 6 hours dealloying, respectively. The length/diameter aspect ratios of the gold ligaments and nanopore channels are approximately between 2 and 3. Thus, the structure of nanoporous gold can be described as a quasi-periodic network made of randomly oriented gold nanowires and hollow nanochannels \cite{Fujita_APL92}. 

In the following, we approximate the structural characteristics of NPG by an ellipsoid: 

\begin{equation}\label{eq_ellipsoid}
\frac{x^2}{a^2}+\frac{y^2}{b^2}+\frac{z^2}{c^2}=1
\end{equation}

\noindent which can be parameterized in polar coordinates as $x=a\sin\alpha\cos\beta$, $y=b\sin\alpha\sin\beta$ and $z=c\cos\alpha$. The local plasmon is excited by the electric field produced by an electron tunneling from the STM tip to a gold particle on the NPG films. The induced dipole moment $\vec{p}$ can be described as:

\begin{equation}\label{eq_dipole}
\left(\begin{matrix}p_x\\p_y\\p_z\end{matrix}\right)=\varepsilon_m\left(\begin{matrix}\alpha_1&0&0\\0&\alpha_2&0\\0&0&\alpha_3\end{matrix}\right)\left(\begin{matrix}E_{0x}\\E_{0y}\\E_{0z}\end{matrix}\right)=\varepsilon_m\boldsymbol{\alpha}\vec{E_0}
\end{equation}

\noindent where $E_{0x}$, $E_{0y}$ and $E_{0z}$ are the components of the electric field $\vec{E_0}$ along the principal axes of the ellipsoid, and $\alpha_i$ $(i=1,2,3)$ represent the polarizability along the principal axes of the polarizability tensor $\boldsymbol\alpha$ of the gold particles. To simplify our calculations, we adopt the system of coordinates $(x', y', z')$ described in the inset of Fig. \ref{Fig3_V_TEM}(c), where the $z'$ axis is defined as the axis connecting the ellipsoid origin $(x = y = z = 0)$ and the point of the electron tunneling defined by the angle $(\theta, \phi)$. The dipole moment is now expressed as:

\begin{equation}\label{eq_pprime}
\vec{p'}=\vec{A}^T\vec{p}=\varepsilon_m(\vec{A}^T\boldsymbol{\alpha}\vec{A})(\vec{A}^T\vec{E_0})
\end{equation}

\noindent where $\vec{A}$ is given by

\begin{equation}\label{eq_A}
\vec{A}=\left(\begin{matrix}\cos\phi&-\cos\theta\sin\phi&-\sin\theta\sin\phi\\ \sin\phi& \cos\theta\cos\phi& \sin\theta\sin\phi \\0& -\sin\theta&\cos\theta\end{matrix}\right)
\end{equation}

\noindent The electric field $\vec{E_0}$ due to the tunneling electron in the $(x, y, z)$ coordinate system is assumed to be given by the electric vector which is normal to the surface of the ellipsoid at the tunneling position with the angle $(\theta, \phi)$ in polar coordinates, and given by the following equation:

\begin{equation}\label{eq_E0}
\vec{E_0}=\left(\begin{matrix}\sin\theta_1\cos\phi_1\\ \sin\theta_1\sin\phi_1\\ \cos\theta_1\end{matrix}\right)
\end{equation}

\noindent where the direction of the electric field is given by the angle $(\theta_1, \phi_1)$ in polar coordinates, which is related to the tip position with the angle $(\theta, \phi)$ and the position of the ellipsoid by the parametric angle $(\alpha, \beta)$. From the ellipsoid geometry, the relationships between the angles $(\theta, \phi)$, $(\theta_1, \phi_1)$ and $(\alpha, \beta)$ are given by

\begin{eqnarray}\label{eq_relations1}
\tan\alpha=\frac{c\tan\theta}{\sqrt{(a^2\cos^2\beta+b^2\sin^2\beta)}}, \phi=\beta\\
\tan^2\theta_1=\tan^2\alpha\left(\frac{c^2}{a^2}\cos^2\beta+\frac{c^2}{b^2}\sin^2\beta\right), \notag\\
\tan\phi_1=\frac{a}{b}\tan\beta
\end{eqnarray}

The electric field in the $(x', y', z)$ coordinate system is given by $\vec{A}^T\vec{E}_0$. In this coordinate system, the dipole moment is given by:

\begin{equation}\label{eq_dipole_p}
\left(\begin{matrix}p'_x\\p'_y\\p'_z\end{matrix}\right)=\varepsilon_m\boldsymbol{\alpha'}\vec{E'}_0=\varepsilon_m(\vec{A}^T\boldsymbol{\alpha}\vec{A})\vec{A}^T\left(\begin{matrix}\sin\theta_1\cos\phi_1\\ \sin\theta_1\sin\phi_1\\ \cos\theta_1\end{matrix}\right)
\end{equation}

\noindent The local plasmon is strongly excited when the above dipole moment has its maximum related to the energy dependent polarizability. The polarizability of the principal axes for the ellipsoid particle are 

\begin{equation}\label{eq_alpha_j}
\alpha_j=4\pi abc\frac{\varepsilon-\varepsilon_0}{3\varepsilon_0+3L_j(\varepsilon-\varepsilon_0)},
\end{equation}

\noindent where

\begin{eqnarray}\label{eq_L_j}
L_1=\frac{abc}{2}\int_0^{\infty}\frac{dq}{(a^2+q)f(q)}, \notag\\
L_2=\frac{abc}{2}\int_0^{\infty}\frac{dq}{(b^2+q)f(q)},L_3=1-L_1-L_2 
\end{eqnarray}

\noindent and

\begin{equation}\label{eq_f}
f(q)=\sqrt{(q+a^2)(q+b^2)(q+c^2)}
\end{equation}

\noindent In the calculation, the complex dielectric constant $\varepsilon$ of the gold particles is taken from experimental data. By using the above equations, we obtain the local plasmon energy from the energy giving the pole of the polarization $\vec{\alpha'}\vec{E'}_0=\varepsilon_0(\vec{A}^T\boldsymbol{\alpha}\vec{A})\vec{A}^T\vec{E}_0$, with use of the energy-dependent dielectric constant of gold for various geometric shapes by changing $a$, $b$ and $c$ at various STM tip positions above the particles. The intensity of the energy loss peak in IETS is proportional to $|{\boldsymbol{\alpha'}\vec{E'}_0}|^2$.

Fig. \ref{Fig3_V_TEM}(c) compares the calculated results for the local plasmon energy as a function of the aspect ratio $c/a$ of the ellipsoid (with $a = b$) for 10 nm and 35 nm pore sizes. In the case of the 10 nm pore size, the calculated energy ranges from 1.2 to 2.5 eV when the aspect ratio varies from 1 to 10, which is consistent with the experimental IETS spectra. As with the experimental results, this range is narrowed and extends from 1.9 to 2.5 eV for the 35 nm pore size, indicating that the local plasmon energy for smaller rod diameters is more sensitive to the $c/a$ structural ratio. The films grown by our group have equivalent ligament and pore sizes, but we caution that this may depend on the dealloying procedure. Other groups may have different structures. In our investigation, it is thus more natural and universal to consider the aspect ratio of the nanorods, as done in Fig. \ref{Fig3_V_TEM}(c), and to compare with parameters determined experimentally by electron tomography analysis. Our model provides a direct way to predict the plasmonic properties of designed materials that does not depend on the dealloying procedure.

According to our model, the energy of the local plasmons depends critically on the shape of the local structures as well as on their orientation with respect to the STM tip position. Therefore, the variation of the IETS spectra and the statistical distribution of the plasmon energy should be closely related to the distribution of the ligament shapes in the films in Figs. \ref{Fig3_V_TEM}(a) and \ref{Fig3_V_TEM}(b). Following this assumption, we transform the abscissa scale (length of ellipsoid) into the energy scale of the local plasmon of the corresponding ellipsoid. The results, displayed in Fig. \ref{Fig3_V_TEM}(d), show distributions centered around 1.7 eV and 2.4 eV for the 10 nm and 35 nm pore size films, respectively, in good agreement with the IETS measurements [see Fig. \ref{Fig2_V_IETS}(c)]. We point out that the calculated distributions are asymmetric curves. The details of these profiles may be due to the uncertainty of the structural distributions. In any case, our calculations capture accurately the origin of the main behavior of the IETS spectra of the NPG films, and demonstrate how the local plasmon evolves with the local geometry of the gold particles.

We note that in the ultraviolet-visible absorption spectra of NPG, the resonant peak from localized surface plasmon resonance excitations red-shifts with increasing the pore size (\emph{i.e}. ligament diameter for our samples), which is opposite to our IETS results \cite{XY_Lang_APL98}. The origin of this difference is unclear but is possibly related to hot electrons production from plasmon decay \cite{Mukherjee_Nano_Lett13}.

\section{Conclusions}

Our experimental and theoretical results suppose that NPG can be considered as interconnected short and curved Au nanorods. The localized electronic properties are investigated by using an UHV-STM at 4.2 K. Second derivative STS shows the plasmon peaks of the NPG films excited by inelastic tunneling electrons. For the film with smaller pore size, more abundant and sharper peaks frequently appear in the low energy range. We propose that the gold nanorod model is appropriate for nanoporous gold studies at the nanometer-scale. This assumption is confirmed by the electron tomography analysis, showing that the length/diameter aspect ratios of the gold nanowires in the NPG films are approximately between 2 and 3. Theoretical calculations of NPG with ellipsoid-like shape, which indicate how the local plasmon evolves with the local geometry of the gold particles, are also in good agreement with the IETS results. 

\vspace{1cm}
\begin{acknowledgments}
H. W. Liu is grateful to C. G. Tao for useful discussions. This work was sponsored by Grand-in-Aid for Scientific Research (B) of JSPS, World Premier International Research Center (WPI) Initiative of MEXT, Sekisui research fund of JST-PRESTO, and JST-CREST of Japan, and by the National Science Foundation and Ministry of Education of China.
\end{acknowledgments}

\bibliography{biblio_HW}

\end{document}